\documentclass[10pt]{iopart}
\usepackage{graphicx}
\begin{document}

\title[Least-squares analysis of clock frequency comparison data]{Least-squares analysis of clock frequency comparison data to deduce optimized frequency and frequency ratio values}

\author{H. S. Margolis and P. Gill}

\address{National Physical Laboratory, Teddington, Middlesex TW11 0LW, UK}
\ead{helen.margolis@npl.co.uk}
\begin{abstract}
A method is presented for analysing over-determined sets of clock frequency comparison data involving standards based on a number of different reference transitions. This least-squares adjustment procedure, which is based on the method used by CODATA to derive a self-consistent set of values for the fundamental physical constants, can be used to derive optimized values for the frequency ratios of all possible pairs of reference transitions. It is demonstrated to reproduce the frequency values recommended by the International Committee for Weights and Measures, when using the same input data used to derive those values. The effects of including more recently published data in the evaluation is discussed and the importance of accounting for correlations between the input data is emphasised.
\end{abstract}

%\submitto{\MET}
%\maketitle

\section{Introduction}

The most advanced optical frequency standards have now reached levels of stability and accuracy~\cite{chou2010,madej2012,letargat2013,hinkley2013,bloom2014,barwood2014,falke2014,godun2014,huntemann2014,ushijima2014,nicholson2014} that significantly surpass the performance of the best caesium primary standards~\cite{levi2006,weyers2012,guena2012,szymaniec2014,heavner2014}, raising the possibility of a future redefinition of the SI second~\cite{gill2011}. As a first step towards preparing for a redefinition the International Committee for Weights and Measures (CIPM), following a recommendation of the Consultative Committee for Time and Frequency (CCTF), introduced the concept of secondary representations of the second that may be used to realize the second in parallel to the caesium primary standard. Seven optical frequency standards (and one microwave frequency standard) may currently be used as secondary representations of the second, with recommended frequencies and uncertainties being assigned by the Frequency Standards Working Group (WGFS) of the CCTF and Consultative Committee for Length (CCL). These recommended frequency values are periodically updated and published on the website of the International Bureau of Weights and Measures~\cite{BIPM_MeP}.

With a single exception, the data considered so far by the WGFS comes from absolute frequency measurements made relative to caesium fountain primary frequency standards. However future information about the reproducibility of the optical standards will come mainly from direct optical frequency ratio measurements~\cite{rosenband2008,matsubara2012,akamatsu2014b,godun2014}. For example, within the EMRP-funded International Timescales with Optical Clocks (ITOC) project~\cite{margolis2013}, a coordinated comparison programme will lead to a set of frequency ratio measurements between high accuracy optical frequency standards being developed within European NMIs, as well as a comprehensive set of absolute frequency measurements with uncertainties at the limit set by caesium primary standards. This set of measurements will be over-determined, in the sense that it will be possible to deduce some of the frequency ratios from several different experiments. For example, a particular frequency ratio $\nu_a/\nu_b$ could be measured directly or it could be determined indirectly by combining two or more other frequency ratio measurements (e.g. $\nu_a/\nu_b = (\nu_a/\nu_c)(\nu_c/\nu_b)$ or $\nu_a/\nu_b = (\nu_a/\nu_d)(\nu_d/\nu_c)(\nu_c/\nu_b)$). If the indirect determinations of this frequency ratio have comparable uncertainties to the direct determination then all of them must be considered in deriving a ``best'' value for the frequency ratio. These multiple routes to deriving each frequency ratio value will complicate the task of the WGFS because it will no longer be possible to treat each optical frequency standard in isolation when considering the available data.

Here we describe a possible approach to analyzing over-determined sets of frequency comparison data to deduce optimized values for the frequency ratios between each contributing standard. The paper is organised as follows. In section~\ref{sec:procedure} we describe the analysis procedure, which follows a least-squares approach. The tests we have carried out to ensure that the algorithms have been correctly implemented in our software are described in section~\ref{sec:tests}. In section~\ref{sec:newdata} we consider the body of frequency comparison data presently available in the published literature, and discuss how the CIPM recommended frequency values might change if recent measurements were included in the evaulation. However this analysis neglects correlations between the individual measurements, and in section~\ref{sec:correlations} we discuss how this affects the results obtained. Finally, section~\ref{sec:conclusion} contains some conclusions and perspectives that may be relevant for future discussions within the WGFS.

\section{Analysis procedure}
\label{sec:procedure}

\begin{figure*}
\includegraphics[width=\columnwidth]{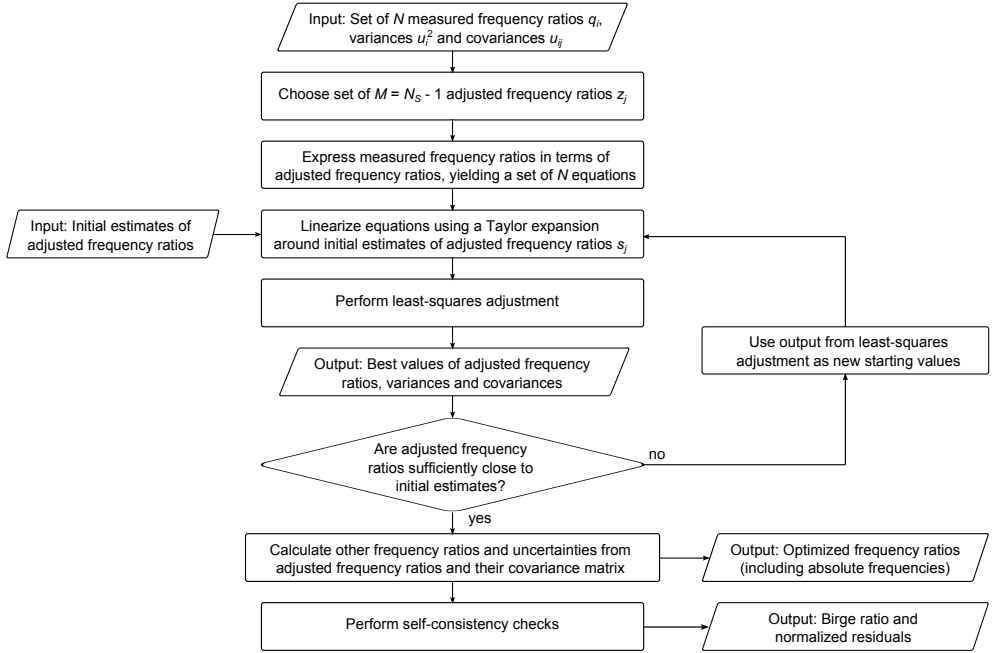}
\caption{Analysis procedure used to derive a self-consistent set of optimized frequency ratio values from a set of frequency comparison experiments involving standards based on $N_{\rm S}$ different reference transitions.}
\label{fig:flowchart}
\end{figure*}

To derive a self-consistent set of optimised frequency ratio values from a set of clock frequency comparison experiments we use a least-squares adjustment procedure. This is based on the approach used by the Committee on Data for Science and Technology (CODATA) to provide a self-consistent set of internationally recommended values of the fundamental physical constants~\cite{mohr2000}. Our method is illustrated in figure~\ref{fig:flowchart}.

Suppose that the frequency standards involved in the comparison experiments are based on $N_{\rm S}$ different reference transitions, with frequencies $\nu_k$ ($k=1,2,\ldots , N_{\rm S}$). For example $\nu_1$ could be the 5s$^2\,^1$S$_0$--5s5p$\,^3$P$_0$ transition in $^{87}$Sr at 698~nm, $\nu_2$ the 6s$\,^2$S$_{1/2}$--4f$^{13}$6s$^2\,^2$F$_{7/2}$ transition in $^{171}$Yb$^+$ at 467~nm, $\nu_3$ the 9.2~GHz 6s$\,^2$S$_{1/2} (F=3)$--6s$\,^2$S$_{1/2} (F=4)$ ground-state hyperfine transition in $^{133}$Cs, and so on. The set of frequency comparison experiments yields a set of $N$ measured values $q_i$ of various quantities, which may be optical frequency ratios, microwave frequency ratios, or optical-microwave frequency ratios. These measured values $q_i$, together with their standard uncertainties $u_i$, variances $u_i^2=u_{ii}$ and covariances $u_{ij}$ (where $u_{ji}=u_{ij}$), form the input data to the least-squares adjustment.

As the first step, we choose a set of $M=N_{\rm S}-1$ adjusted frequency ratios $z_j$ (where $M<N$) given by
\begin{equation}
z_j = \frac{\nu_j}{\nu_{j+1}}\,,
\end{equation}
where $j=1,2,\ldots , N_S -1$.
These adjusted frequency ratios are the variables in the least-squares adjustment and are equivalent to the adjusted constants in the CODATA analysis of the fundamental physical constants. Our choice of $z_j$ meets the necessary condition that no adjusted frequency ratio may be expressed as a function of the others.

The quantities $q_i$ that are measured in the frequency comparison experiments are next expressed as a function $f_i$ of one or more of these adjusted frequency ratios $z_j$ by the set of $N$ equations
\begin{equation}
q_i\doteq f_i(z_1, z_2, \ldots , z_M) 
\label{eq:obs_eqns}
\end{equation}
where $i=1,2,\ldots ,N$.
Here the symbol $\doteq$ is used to indicate that in general the left and right hand sides of the equation are not exactly equal, because the set of equations is overdetermined. For example, the first observed quantity $q_1$ might be 
$\nu_2/\nu_5$, which can be expressed as $z_2 z_3 z_4$, while the second observed quantity $q_2$ might be either another measurement of $\nu_2/\nu_5$ or a measurement of a different ratio such as $\nu_1/\nu_6$.

The equations~(\ref{eq:obs_eqns}) are, in most cases, nonlinear. Prior to the least-squares adjustment, they are therefore
linearized using a Taylor expansion around starting values $s_j$ (initial estimates of the adjusted frequency ratios):
\begin{eqnarray}
q_i & \doteq & f_i(s_1,s_2,\ldots ,s_M) \nonumber\\
& & + \sum^M_{j=1}\frac{\partial f_i(s_1,s_2,\ldots ,s_M)}{\partial s_j}(z_j-s_j) + \cdots \,. \label{eq:lin_obs_eqns}
\end{eqnarray}
This enables linear matrix methods to be applied. Defining new variables 
\begin{equation}
y_i = q_i - f_i(s_1,s_2,\ldots ,s_M)
\end{equation}
and
\begin{equation}
x_j = z_j - s_j\,,
\end{equation}
the linearized equations~(\ref{eq:lin_obs_eqns}) can be rewritten in the form
\begin{equation}
y_i \doteq \sum^M_{j=1} a_{ij}x_j
\end{equation}
where
\begin{equation}
a_{ij} = \frac{\partial f_i(s_1,s_2,\ldots ,s_M)}{\partial s_j}\,.
\end{equation}
In matrix notation these equations become
\begin{equation}
\bi{Y} \doteq \bi{A X}\,,
\label{eq:matrix_form}
\end{equation}
where $\bi{Y}$ is a column matrix with $N$ elements, $\bi{A}$ is a rectangular matrix with $N$ rows and $M$ columns, and $\bi{X}$ is a column matrix with $M$ elements. In the same way, we consider the measured frequency ratios $q_i$, the functions $f_i$ and the adjusted frequency ratios $z_j$ and their initial estimates $s_j$ to be elements of matrices $\bi{Q}$, $\bi{F}$, $\bi{Z}$ and $\bi{S}$ respectively. 

To obtain the best estimate of $\bi{X}$, and hence of the adjusted frequency ratios $\bi{Z}$, a least-squares adjustment is performed, minimizing the product 
\begin{equation}
S = (\bi{Y}-\bi{AX})^T \bi{V}^{-1}(\bi{Y}-\bi{AX}) \label{eq:least_squares}
\end{equation}
with respect to $\bi{X}$. Here $\bi{V}=\rm{cov}\bi{Y}$ is the $N\times N$ covariance matrix of the input data, with elements $u_{ij}$. 
The solution $\hat{\bi{X}}$ is used to calculate the best estimate $\hat{\bi{Z}}$ of the adjusted frequency ratios according to
\begin{equation}
\hat{\bi{Z}} = \bi{S} + \hat{\bi{X}}\,.
\end{equation}

Although the solution of the linear approximation~(\ref{eq:lin_obs_eqns}) does not provide an exact solution of the nonlinear equations~(\ref{eq:obs_eqns}), the values of the adjusted frequency ratios $\hat{\bi{Z}}$ obtained from the least-squares adjustment will be an improvement over the starting values $\bi{S}$. To obtain more precise values, these improved values of the adjusted frequency ratios are used as starting values for a new linear approximation and a second least-squares adjustment is performed. This procedure is repeated until the new values of the adjusted frequency ratios obtained from the least-squares adjustment differ from the starting values for that iteration by a sufficiently small fraction of their uncertainties. The number of iterations required to satisfy this condition will depend on how close the starting values $s_j$ are to the values of $\hat{z}_j$ calculated in the final iteration.
Once convergence has been achieved, the best estimate $\hat{\bi{Q}}$ of the measured quantities $\bi{Q}$ can be calculated from the final solution $\hat{\bi{X}}$:
\begin{equation}
\hat{\bi{Q}} = \bi{F} + \bi{A}\hat{\bi{X}}\,.
\end{equation}

In general, the values of the adjusted frequency ratios obtained from the least-squares adjustment will be correlated. The covariance matrix 
${\rm cov}(\hat{\bi{Z}})={\rm cov}(\hat{\bi{X}})$, whose elements $u(\hat{z}_i,\hat{z}_j)$ are the covariances of the adjusted frequency ratios, can be used to evaluate the uncertainty in other frequency ratios calculated from two or more of the adjusted frequency ratios. In general, there will be a total of $K$ possible frequency ratios $\hat{r}_i$ which may be expressed in terms of the adjusted frequency ratios $\hat{z}_j$:
\begin{equation}
\hat{r}_i (\hat{z}_1, \hat{z}_2, \ldots , \hat{z}_M) \label{eq:other_ratios}
\end{equation}
where $i=1,2,\ldots ,K$, including expressions of the form $\hat{r}_i=\hat{z}_j$.
According to the standard formula for the propagation of uncertainty, the covariances of these optimized frequency ratios are given by
\begin{equation}
u(\hat{r}_k,\hat{r}_l) = \sum^M_{i,j=1}\frac{\partial\hat{r}_k}{\partial\hat{z}_i}\frac{\partial\hat{r}_l}{\partial\hat{z}_j}u(\hat{z}_i,\hat{z}_j) \,. \label{eq:uncert_calc}
\end{equation}
If $l=k$, then equation~(\ref{eq:uncert_calc}) gives the variances $u^2(\hat{r}_k)=u(\hat{r}_k,\hat{r}_k)$\,.

Self-consistency checks on the body of data provide verification of the uncertainty evaluations for each individual frequency standard and enable any issues with individual frequency standards to be identified.
To obtain a measure of the consistency of the input data, the Birge ratio
\begin{equation}
R_{\rm B} = \left( \frac{\chi^2}{N-M} \right)^{1/2}\,,
\end{equation}
is computed. Here
\begin{equation}
\chi^2 = (\bi{Q} - \hat{\bi{Q}})^T \bi{V}^{-1} (\bi{Q} - \hat{\bi{Q}})\,,
\end{equation}
is the minimum value of $S$ as given by equation~(\ref{eq:least_squares}), evaluated in the final iteration of the least-squares adjustment. A Birge ratio significantly larger than one suggests that the input data are inconsistent. Similarly, a normalized residual
\begin{equation}
\rho_i = \frac{q_i - \hat{q}_i}{u_i}
\end{equation}
significantly larger than one for a particular measured frequency ratio $q_i$ suggests that the measurement is inconsistent with the other data.

These algorithms have been implemented in Matlab, with the least-squares solution to equation~(\ref{eq:matrix_form}) being found using the routine {\tt lscov()}. Due to the extremely high accuracy with which frequency comparisons can be performed, numerical calculations must be performed with a precision of more than 18 significant figures. This is achieved using routines designed for high precision floating point arithmetic~\cite{derrico2012}.

\section{Tests of the software algorithms}
\label{sec:tests}

Several simulated test data sets were generated for testing the analysis software. These were generated in such a way that the software, if functioning correctly, would generate certain known values for the absolute frequencies of each reference transition. Since the software does not calculate these frequencies or their uncertainties directly, but instead calculates them from combinations of the adjusted frequency ratios and the covariance matrix determined from the least-squares adjustment, this approach is considered to provide a good check that the adjusted frequency ratios are also being calculated correctly.

One key test of the software algorithms employed in the analysis procedure is whether they are able to reproduce the CIPM recommended frequency values, when using the same input data employed by the CCL-CCTF WGFS. To check this, the set of $N_S=15$ frequency standards listed in table~\ref{tab:standards} was considered.

\begin{table}
\caption{\label{tab:standards}Frequency standards considered in the tests of the least-squares adjustment procedure.}
\begin{indented}
\item[]\begin{tabular}{@{}ll}
\br
Atom/ion & Reference transition \\
\mr
$^{115}$In$^{+}$ & 5s$^2\,^1$S$_0$--5s5p$\,^3$P$_0$ \\
$^1$H & 1s$\,^2$S$_{1/2}$--2s$\,^2$S$_{1/2}$ \\
$^{199}$Hg & 6s$^2\,^1$S$_0$--6s6p$\,^3$P$_0$ \\
$^{27}$Al$^+$ & 3s$^2\,^1$S$_0$--3s3p$\,^3$P$_0$ \\
$^{199}$Hg$^+$ & 5d$^{10}$6s$\,^2$S$_{1/2}$--5d$^9$6s$^2\,^2$D$_{5/2}$ \\
$^{171}$Yb$^+$ & 6s$\,^2$S$_{1/2}$--5d$\,^2$D$_{3/2}$ \\
$^{171}$Yb$^+$ & 6s$\,^2$S$_{1/2}$--4f$^{13}$6s$^2\,^2$F$_{7/2}$ \\
$^{171}$Yb & 6s$^2\,^1$S$_0$--6s6p$\,^3$P$_0$ \\
$^{40}$Ca & 4s$^2\,^1$S$_0$--4s4p$\,^3$P$_1$ \\
$^{88}$Sr$^+$ & 5s$\,^2$S$_{1/2}$--4d$\,^2$D$_{5/2}$ \\
$^{88}$Sr & 5s$^2\,^1$S$_0$--5s5p$\,^3$P$_0$ \\
$^{87}$Sr & 5s$^2\,^1$S$_0$--5s5p$\,^3$P$_0$ \\
$^{40}$Ca$^+$ & 4s$\,^2$S$_{1/2}$--3d$\,^2$D$_{5/2}$ \\
$^{87}$Rb & 5s$\,^2$S$_{1/2} (F=1)$--5s$\,^2$S$_{1/2} (F=2)$ \\
$^{133}$Cs & 6s$\,^2$S$_{1/2} (F=3)$--6s$\,^2$S$_{1/2} (F=4)$ \\
\br
\end{tabular}
\end{indented}
\end{table}

\begin{figure}
\hspace{2cm}
\includegraphics[width=0.8\columnwidth]{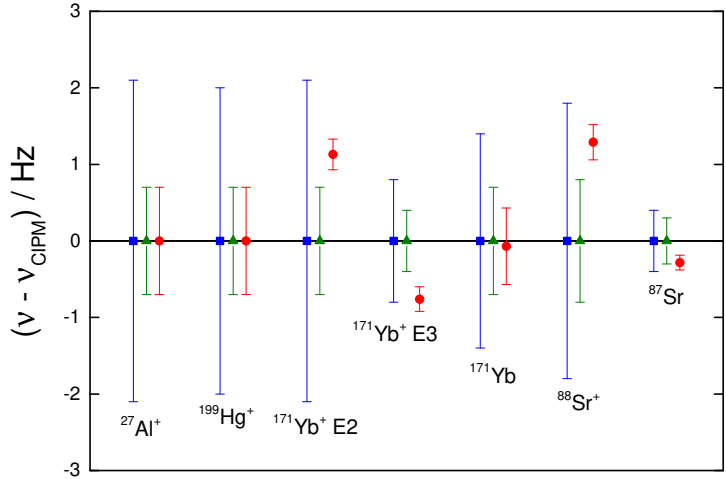}
\caption{Frequency values obtained for the seven optical secondary representations of the second, calculated using the same input data used by the CCL-CCTF WGFS (green triangles) and with new data included in the analysis (red circles). The present CIPM recommended frequency values are also shown (blue squares).}
\label{fig:secrep_results}
\end{figure}

The results obtained for the seven optical secondary representations of the second are shown in figure~\ref{fig:secrep_results}. All these frequency values agree with the CIPM values, but the uncertainties determined from the least-squares adjustment procedure are smaller than the uncertainties assigned by the CIPM, sometimes by a factor of two or three. The explanation for this is that the WGFS takes a conservative approach to estimating uncertainties, typically multiplying the relative standard uncertainty on the weighted mean of a set of frequency values by a factor of two or three to reflect the fact that the measurements originate from only a few independent research groups (or in some cases a single research group). Our analysis procedures yield uncertainties equivalent to the relative standard uncertainty.

The frequency values we obtain for the other standards listed in table~\ref{tab:standards} are in similarly good agreement with the CIPM values, with the sole exception of the $^{40}$Ca optical frequency standard. The reason for this discrepancy is that in this case the WGFS departed from its normal procedure of taking a weighted mean of the available frequency measurements. Instead, bearing in mind a significant discrepancy between two independent frequency measurements, an unweighted mean of the two values was used. Our analysis procedures, on the other hand, are equivalent to taking a weighted mean of the two values.

\section{Inclusion of new frequency comparison data}
\label{sec:newdata}

Since the last review of available frequency data by the WGFS, which was performed in September 2012, a number of new frequency comparison results have been reported. These include ten new absolute frequency measurements (one of the 1S--2S transition in $^1$H~\cite{matveev2013}, two each of the E2~\cite{tamm2014,godun2014} and E3~\cite{godun2014,huntemann2014} reference transitions in $^{171}$Yb$^+$, one each of the reference transitions in $^{171}$Yb~\cite{park2013} and $^{88}$Sr$^+$~\cite{barwood2014} and three of the reference transition in $^{87}$Sr~\cite{letargat2013,akamatsu2014a,falke2014}). However three optical frequency ratios have also been measured directly. These are the ratios between the reference transitions in $^{40}$Ca$^+$ and $^{87}$Sr~\cite{matsubara2012}, between the reference transitions in $^{171}$Yb and $^{87}$Sr~\cite{akamatsu2014b}, and between the E2 and E3 reference transitions in $^{171}$Yb$^+$~\cite{godun2014}.

The effects on the optimized frequency values obtained for the optical secondary representations of the second of including this additional data in the analysis (but at this stage ignoring any correlations between the input data) are shown in figure~\ref{fig:secrep_results}. Significant changes are observed for some of the optimized frequency values, most notably the $^{171}$Yb$^+$ and $^{88}$Sr$^+$ optical frequency standards, but also for $^{87}$Sr. This illustrates that the conservative approach adopted by the WGFS in assigning uncertainties to the recommended frequency values may indeed be prudent.

The $^{87}$Sr data is, however, coupled to the data for $^{40}$Ca$^+$ by the optical frequency ratio measurement between these two reference transitions~\cite{matsubara2012}, and the presently available body of data for $^{40}$Ca$^+$ is not internally self-consistent. To assess how much effect this coupling has on the optimized frequency value for the $^{87}$Sr reference transition, the $^{40}$Ca$^+$/$^{87}$Sr frequency ratio measurement was removed from the input data and the least-squares adjustment repeated. The difference from the current CIPM recommended frequency value for $^{87}$Sr was found to reduce only slightly from 0.28~Hz to 0.25~Hz.

\section{Importance of correlations}
\label{sec:correlations}

The above analysis neglects any correlations between the measured frequency ratio values used as input to the least-squares adjustment. In reality, there will be correlations among the input data that should be accounted for. For example, at NPL, the absolute frequencies of the E2 and E3 reference transitions in $^{171}$Yb$^+$ and the direct optical frequency ratio between them were recently all determined during a single measurement campaign, with substantial periods of overlap in the data-taking periods~\cite{godun2014}. This means that there are significant correlations between these three values. As frequency comparison experiments involving multiple optical frequency standards and even multiple laboratories become more frequent, such correlations will become more frequent and more significant. However the covariances and corresponding correlation coefficients between different frequency ratio measurements are not normally reported in the literature, even for values obtained in the same laboratory. To evaluate these covariances additional information would typically need to be obtained from each research group involved, just as in the CODATA least-squares adjustment of the fundamental physical constants~\cite{mohr2000}.

\begin{figure}
\centering
\hspace{2cm}
\includegraphics[width=0.7\columnwidth]{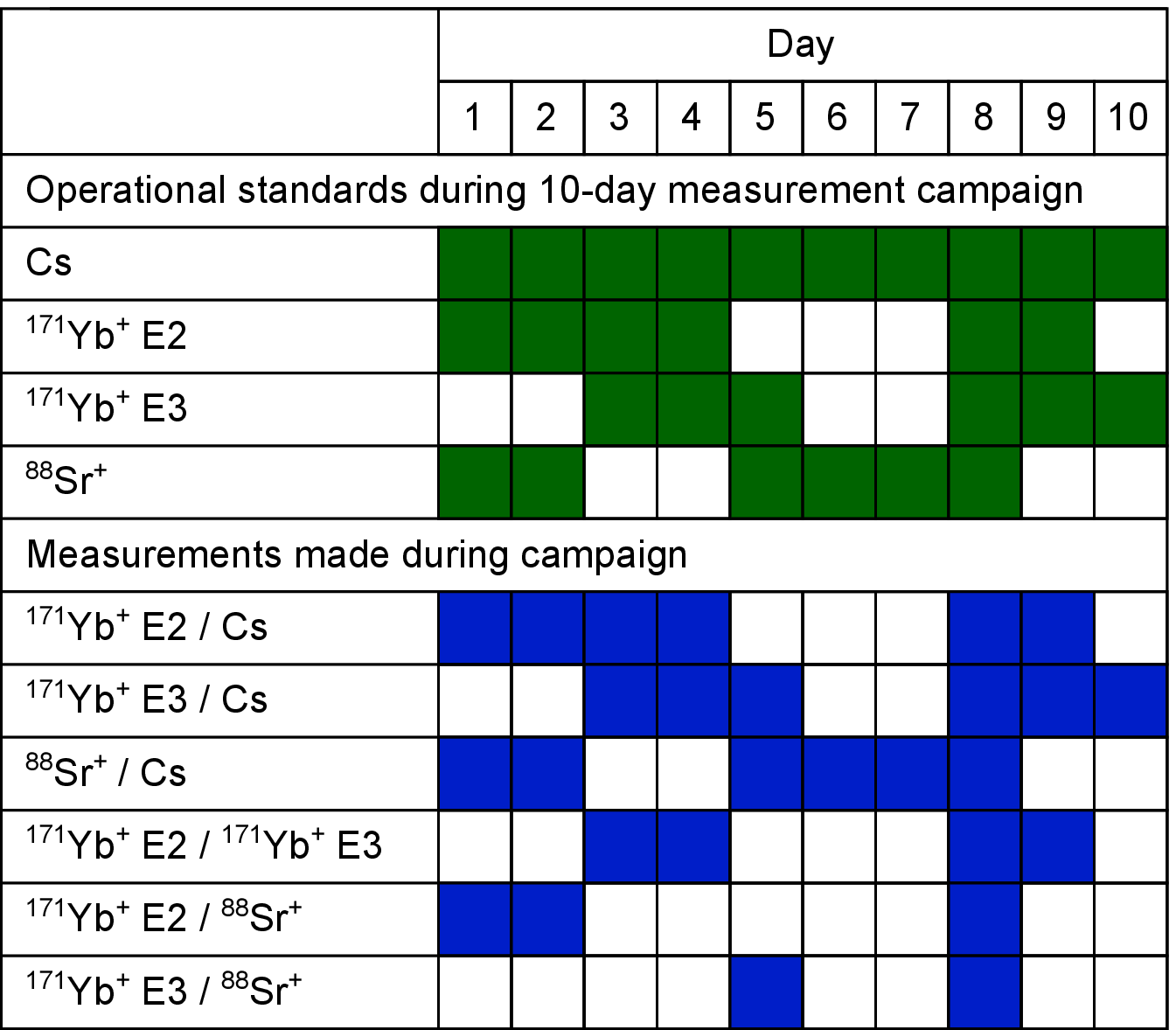}
\caption{Hypothetical 10-day measurement campaign involving four different frequency standards. The top half of the figure shows which standards are operating during each day of the campaign, whilst the lower half shows which frequency ratios can be determined from the data taken each day.}
\label{fig:campaign}
\end{figure}

To illustrate the importance of properly accounting for correlations, we consider the hypothetical 10-day measurement campaign illustrated in figure~\ref{fig:campaign}. This involves a caesium primary frequency standard, which we assume operates 100\% of the time, and three optical frequency standards, which we assume each run for 60\% of the time, with some periods of overlap. Six different frequency ratios can be determined, each from different periods of the campaign. For these six frequency ratios, there are twelve non-zero correlation coefficients.

Correlations arise from both statistical and systematic uncertainties. For example, the absolute frequency measurement of the E2 transition in $^{171}$Yb$^+$ and the absolute frequency measurement of the reference transition in $^{88}$Sr$^+$ are correlated because part of the caesium primary frequency standard data is common to the two. Assuming initially that all other sources of uncertainty are negligible compared to the statistical uncertainty associated with the caesium standard, then since the $^{171}$Yb$^+$ E2 standard runs for a total period $T_A=6$~days and the $^{88}$Sr$^+$ standard runs for a total period $T_B=6$~days, with a period of overlap $T_{\rm overlap}=3$~days, we calculate a correlation coefficient of $\sqrt{T_{\rm overlap}^2/(T_A T_B)} = 0.5$. However in practice for a measurement of this duration, our initial assumption is not a good one and other contributions to the uncertainty must be considered. For this particular example, there is an additional source of correlation because the systematic uncertainty of the caesium fountain is also common to the two measurements. In a similar way, correlations between the absolute frequency measurement of the E3 transition in $^{171}$Yb$^+$ and the $^{171}$Yb$^+ \rm{E2} / ^{171}$Yb$^+ {\rm E3}$ frequency ratio measurement arise from both the statistical and the systematic uncertainty of the $^{171}$Yb$^+$ E3 standard. 
If we use the present stabilities and systematic uncertainties of NPL's frequency standards~\cite{barwood2014,godun2014,szymaniec2014} to estimate the correlation coefficients for this hypothetical measurement campaign, we find that the values of the twelve correlation coefficients range from $-0.102$ to $0.95$. The largest correlation coefficient is for the Yb$^+$ E2 / Yb$^+$ E3 and Yb$^+$ E2 / $^{88}$Sr$^+$ frequency ratios, since the uncertainties of both measurements would be dominated by the systematic uncertainty of the Yb$^+$ E2 standard at its current state of development.

\begin{figure}
\hspace{2cm}
\includegraphics[width=0.8\columnwidth]{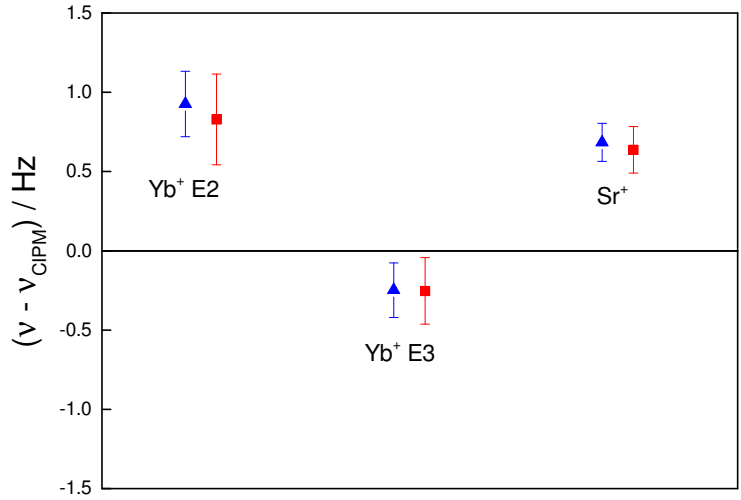}
\caption{Effect of correlations on the output from the least-squares adjustment for the hypothetical measurement campaign illustrated in figure~\ref{fig:campaign}. Blue triangles show the absolute frequency values obtained when correlations are neglected, whilst red squares show the corresponding values obtained when correlations are included.}
\label{fig:correlations}
\end{figure}

For arbitrarily-selected values of the measured frequency ratios resulting from this hypothetical 10-day measurement campaign, the effect of correlations on the optimized frequency ratios and absolute frequencies can be determined. As illustrated in figure~\ref{fig:correlations}, neglecting correlations leads to too much weight being given to these measurements in the least-squares adjustment, resulting in biased frequency values and underestimated uncertainties. For some planned measurement campaigns, even stronger correlations could potentially arise, resulting in more significant biases if these correlations were neglected in the analysis. This demonstrates the importance of gathering information about the correlations between the input data, both for intra-laboratory frequency comparisons and for inter-laboratory frequency comparisons.

\section{Conclusion}
\label{sec:conclusion}

In summary, we have described an analysis procedure that can be applied to determine a self-consistent set of frequency ratios between high accuracy frequency standards (both optical and microwave) based on all experimental data available at any particular time, and including correlations among the data. Currently the matrix of frequency comparison data is rather sparsely populated, but as the number of frequency ratio measurements made without reference to the caesium primary standard increases, our methods and software could be used to provide valuable information about the relative performance of different candidates for an optical redefinition of the SI second. They can also be used to determine optimized values and uncertainties for the absolute frequencies of each optical standard relative to the current definition of the SI second, since these are simply special cases of frequency ratios involving the caesium primary standard. Such absolute frequency values will be required to maximize the potential contribution of optical clocks to international timescales prior to any redefinition.

Our work also identifies the key issues likely to be encountered in assessing an over-determined set of frequency comparison data. Firstly, it will be necessary to identify and critically review all possible input data, with particular attention being paid to the standard uncertainty associated with each measurement. Secondly, the correlations between the input data must be considered, since these can have a significant effect on the frequency ratio values and uncertainties obtained from the least-squares adjustment. However the information reported in the literature is in many cases insufficient to calculate the correlation coefficients, meaning that it will be necessary to seek additional information from the groups that carried out the measurements in order that each input datum is given the appropriate weight in the least-squares adjustment. Finally, it will be necessary to investigate the extent to which each input datum contributes to the determination of the adjusted frequency ratios as well as the effects of omitting inconsistent or inconsequential data as may be deemed appropriate. None of these issues are unique to the problem discussed here; indeed they are common to those that have been faced by the CODATA Task Group on Fundamental Constants for many years. As such they are likely to be highly relevant to future discussions within the CCL-CCTF Frequency Standards Working Group.

\ack

This work was performed within the ITOC project as part of the European Metrology Research Programme (EMRP).
The EMRP is jointly funded by the EMRP participating countries within EURAMET and the European Union.
The authors gratefully acknowledge helpful discussions with partners in the ITOC project consortium and with members of the CCL-CCTF WGFS.

\section*{References}
\bibliographystyle{unsrt}
\bibliography{FreqRatioAnalysis}

\begin{thebibliography}{10}

\bibitem{chou2010}
C.~W. Chou, D.~B. Hume, J.~C.~J. Koelemeij, D.~J. Wineland, and T.~Rosenband.
\newblock Frequency comparison of two high-accuracy {A}l$^+$ optical clocks.
\newblock {\em Phys. Rev. Lett.}, 104:070802, 2010.

\bibitem{madej2012}
A.~A. Madej, P.~Dub{\'e}, Z.~Zhou, J.~E. Bernard, and M.~Gertsvolf.
\newblock $^{88}${S}r$^+$ 445-{TH}z single-ion reference at the $10^{-17}$
  level via control and cancellation of systematic uncertainties and its
  measurement against the {SI} second.
\newblock {\em Phys. Rev. Lett.}, 109:203002, 2012.

\bibitem{letargat2013}
R.~Le Targat, L.~Lorini, Y.~Le Coq, M.~Zawada, J.~Gu{\'e}na, M.~Abgrall,
  M.~Gurov, P.~Rosenbusch, D.~G. Rovera, B.~Nag{\'o}ney, R.~Gartman, P.~G.
  Westergaard, M.~E. Tobar, M.~Lours, G.~Santarelli, A.~Clairon, S.~Bize,
  P.~Laurent, P.~Lemonde, and J.~Lodewyck.
\newblock Experimental realization of an optical second with strontium lattice
  clocks.
\newblock {\em Nat. Commun.}, 4:2109, 2013.

\bibitem{hinkley2013}
N.~Hinkley, J.~A. Sherman, N.~B. Phillips, M.~Schioppo, N.~D. Lemke, K.~Beloy,
  M.~Pizzocaro, C.~W. Oates, and A.~D. Ludlow.
\newblock An atomic clock with $10^{-18}$ instability.
\newblock {\em Science}, 341:1215--1218, 2013.

\bibitem{bloom2014}
B.~J. Bloom, T.~L. Nicholson, J.~R. Williams, S.~L. Campbell, M.~Bishof,
  X.~Zhang, W.~Zhang, S.~L. Bromley, and J.~Ye.
\newblock An optical lattice clock with accuracy and stability at the
  $10^{-18}$ level.
\newblock {\em Nature}, 506:71--75, 2014.

\bibitem{barwood2014}
G.~P. Barwood, G.~Huang, H.~A. Klein, L.~A.~M. Johnson, S.~A. King, H.~S.
  Margolis, K.~Szymaniec, and P.~Gill.
\newblock Agreement between two $^{88}${S}r$^{+}$ optical clocks to 4 parts in
  10$^{17}$.
\newblock {\em Phys. Rev. A}, 89:050501, 2014.

\bibitem{falke2014}
S.~Falke, N.~Lemke, C.~Grebing, B.~Lipphardt, S.~Weyers, V.~Gerginov,
  N.~Huntemann, C.~Hagemann, A.~A-Masoudi, S.~H{\"a}fner, S.~Vogt, U.~Sterr,
  and C.~Lisdat.
\newblock A strontium lattice clock with $3\times 10^{-17}$ inaccuracy and its
  frequency.
\newblock {\em New. J. Phys.}, 16:073023, 2014.

\bibitem{godun2014}
R.~M. Godun, P.~B.~R. Nisbet-Jones, J.~M. Jones, S.~A. King, L.~A.~M. Johnson,
  H.~S. Margolis, K.~Szymaniec, S.~N. Lea, K.~Bongs, and P.~Gill.
\newblock Frequency ratio of two optical clock transitions in
  $^{171}${Y}b$^{+}$ and constraints on the time-variation of fundamental
  constants.
\newblock {\em Phys. Rev. Lett.}, 113:210801, 2014.

\bibitem{huntemann2014}
N.~Huntemann, B.~Lipphardt, C.~Tamm, V.~Gerginov, S.~Weyers, and E.~Peik.
\newblock Improved limit on a temporal variation of $m_p/m_e$ from comparisons
  of {Y}b$^+$ and {C}s atomic clocks.
\newblock {\em Phys. Rev. Lett.}, 113:210802, 2014.

\bibitem{ushijima2014}
I.~Ushijima, M.~Takamoto, M.~Das, T.~Ohkubo, and H.~Katori.
\newblock Cryogenic optical lattice clocks with a relative frequency difference
  of $1\times 10^{-18}$.
\newblock {\em arXiv:1405.4071}, 2014.

\bibitem{nicholson2014}
T.~L. Nicholson, S.~L. Campbell, R.~B. Hutson, G.~E. Marti, B.~J. Bloom, R.~L.
  McNally, W.~Zhang, M.~D. Barrett, M.~S. Safronova, G.~F. Strouse, W.~L. Tew,
  and J.~Ye.
\newblock $2\times 10^{-18}$ uncertainty in an atomic clock.
\newblock {\em arXiv:1412.8261}, 2014.

\bibitem{levi2006}
F.~Levi, D.~Calonico, L.~Lorini, and A.~Godone.
\newblock {IEN}-{C}s{F}1 primary frequency standard at {INRIM}: accuracy
  evaluation and {TAI} calibrations.
\newblock {\em Metrologia}, 43:545--555, 2006.

\bibitem{weyers2012}
S.~Weyers, V.~Gerginov, N.~Nemitz, R.~Li, and K.~Gibble.
\newblock Distributed cavity phase frequency shifts of the caesium fountain
  {PTB}-{C}s{F}2.
\newblock {\em Metrologia}, 49:82--87, 2012.

\bibitem{guena2012}
J.~Gu{\'e}na, M.~Abgrall, D.~Rovera, P.~Laurent, B.~Chupin, M.~Lours,
  G.~Santarelli, P.~Rosenbusch, M.~E. Tobar, R.~Li, K.~Gibble, A.~Clairon, and
  S.~Bize.
\newblock Progress in atomic fountains at {LNE-SYRTE}.
\newblock {\em IEEE Trans. Ultrason. Ferroelectr. Freq. Control}, 59:391--410,
  2012.

\bibitem{szymaniec2014}
K.~Szymaniec, S.~Lea, and K.~Liu.
\newblock An evaluation of the frequency shift caused by collisions with
  background gas in the primary frequency standard {NPL}-{C}s{F}2.
\newblock {\em IEEE Trans. Ultrason. Ferroelectr. Freq. Control}, 61:203--206,
  2014.

\bibitem{heavner2014}
T.~P. Heavner, E.~A. Donley, F.~Levi, G.~Costanzo, T.~E. Parker, J.~H. Shirley,
  N.~Ashby, S.~Barlow, and S.~R. Jefferts.
\newblock First accuracy evaluation of {NIST}-{F}2.
\newblock {\em Metrologia}, 51:174--182, 2014.

\bibitem{gill2011}
P.~Gill.
\newblock When should we change the definition of the second?
\newblock {\em Proc. Roy. Soc. A}, 369:4109--4130, 2011.

\bibitem{BIPM_MeP}
BIPM.
\newblock Recommended values of standard frequencies.
\newblock
  http://www.bipm.org/en/publications/mises-en-pratique/standard-frequencies.html.

\bibitem{rosenband2008}
T.~Rosenband, D.~B. Hume, P.~O. Schmidt, C.~W. Chou, A.~Brusch, L.~Lorini,
  W.~H. Oskay, R.~E. Drullinger, T.~M. Fortier, J.~E. Stalnaker, S.~A. Diddams,
  W.~C. Swann, N.~R. Newbury, W.~M. Itano, D.~J. Wineland, and J.~C. Bergquist.
\newblock Frequency ratio of {A}l$^+$ and {H}g$^+$ single-ion optical clocks;
  metrology at the 17th decimal place.
\newblock {\em Science}, 319:1808--1811, 2008.

\bibitem{matsubara2012}
K.~Matsubara, H.~Hachisu, Y.~Li, S.~Nagano, C.~Locke, A.~Nogami, M.~Kajita,
  K.~Hayasaka, T.~Ido, and M.~Hosokawa.
\newblock Direct comparison of a {C}a$^+$ single-ion clock against a {S}r
  lattice clock to verify the absolute frequency measurement.
\newblock {\em Opt. Express}, 20:22034--22041, 2012.

\bibitem{akamatsu2014b}
D.~Akamatsu, M.~Yasuda, H.~Inaba, K.~Hosaka, T.~Tanabe, A.~Onae, and F.-L.
  Hong.
\newblock Frequency ratio measurement of $^{171}${Y}b and $^{87}${S}r optical
  lattice clocks.
\newblock {\em Opt. Express}, 22:7898--7905, 2014.

\bibitem{margolis2013}
H.~S. Margolis, R.~M. Godun, P.~Gill, L.~A.~M. Johnson, S.~L. Shemar, P.~B.
  Whibberley, D.~Calonico, F.~Levi, L.~Lorini, M.~Pizzocaro, P.~Delva, S.~Bize,
  J.~Achkar, H.~Denker, L.~Timmen, C.~Voigt, S.~Falke, D.~Piester, C.~Lisdat,
  U.~Sterr, S.~Vogt, S.~Weyers, J.~Gersl, T.~Lindvall, and M.~Merimaa.
\newblock International timescales with optical clocks ({ITOC}).
\newblock In {\em Proceedings of the 2013 Join European Frequency and Time
  Forum and International Frequency Control Symposium}, pages 908--911. {IEEE},
  2013.

\bibitem{mohr2000}
P.~J. Mohr and B.~N. Taylor.
\newblock {CODATA} recommended values of the fundamental physical constants:
  1998.
\newblock {\em Rev. Mod. Phys.}, 72:351--495, 2000.

\bibitem{derrico2012}
J.~R. D'Errico.
\newblock {HPF} class, 2012.
\newblock Available from the Matlab central file exchange, {\tt
  http://www.mathworks.com/matlabcentral/fileexchange/}.

\bibitem{matveev2013}
A.~Matveev, C.~G. Parthey, K.~Predehl, J.~Alnis, A.~Beyer, R.~Holzwarth,
  T.~Udem, T.~Wilken, N.~Kolachevsky, M.~Abgrall, D.~Rovera, C.~Salomon,
  P.~Laurent, G.~Grosche, O.~Terra, T.~Legero, H.~Schnatz, S.~Weyers,
  B.~Altschul, and T.~W. H{\"a}nsch.
\newblock Precise measurement of the hydrogen 1{S}--2{S} frequency via a 920-km
  fiber link.
\newblock {\em Phys. Rev. Lett.}, 110:230801, 2013.

\bibitem{tamm2014}
C.~Tamm, N.~Huntemann, B.~Lipphardt, V.~Gerginov, N.~Nemitz, M.~Kazda,
  S.~Weyers, and E.~Peik.
\newblock Cs-based optical frequency measurement using cross-linked optical and
  microwave oscillators.
\newblock {\em Phys. Rev. A}, 89:023820, 2014.

\bibitem{park2013}
C.~Y. Park, D.-H. Yu, W.-K. Lee, S.~E. Park, E.~B. Kim, S.~K. Lee, J.~W. Cho,
  T.~H. Yoon, J.~Mun, S.~J. Park, T.~Y. Kwon, and S.-B. Lee.
\newblock Absolute frequency measurement of $^1${S}$_0 ({F}=1/2)$--$^3${P}$_0
  ({F}=1/2)$ transition of $^{171}${Y}b atoms in a one-dimensional optical
  lattice at {KRISS}.
\newblock {\em Metrologia}, 50:119--128, 2013.

\bibitem{akamatsu2014a}
D.~Akamatsu, H.~Inaba, K.~Hosaka, M.~Yasuda, A.~Onae, T.~Suzuyama, M.~Amemiya,
  and F.-L. Hong.
\newblock Spectroscopy and frequency measurement of the $^{87}${S}r clock
  transition by laser linewidth transfer using an optical frequency comb.
\newblock {\em Appl. Phys. Express}, 7:012401, 2014.

\end{thebibliography}

\end{document}